# Network model of human language

Mária Markošová
Dept. of Applied Informatics,
Faculty of Mathematics, Physics and Informatics
Mlynská dolina, Bratislava, Slovakia.

November 14, 2018

## Abstract

*The phenomenon of human language is widely studied from various points of view. It is interesting not only for social scientists, antropologists or philosophers, but also for those, interesting in the network dynamics. In several recent papers word web, or language as a graph has been investigated [1, 2, 3].*

*In this paper I revise recent studies of syntactical word web [1, 4]. I present a model of growing network in which such processes as node addition, edge rewiring and new link creation are taken into account. I argue, that this model is a satisfactory minimal model explaining measured data [1, 5].*

# 1 Introduction

Networks are nowadays very popular to investigate. They are good models for various types of interactions, such as social interactions, professional interactions [6], interactions in biology [7], interaction as communication [8, 9] to which belong also interactions of people through language [1, 2, 4, 5]. Networks are also an interesting objects to study theoretically, because their properties are strongly influenced by the network history and dynamics. Network can grow with time by node addition, the nodes can extinct, too. Several questions has been asked about the details of the net dynamics. For example, how the dynamics influences the overall network structure [10, 11, 13], or what is the dynamics governing real networks [12]?

Network is a collection of nodes interacting through edges. Binary undirected networks are the simplest ones; the edge between two nodes either exists or not. Networks are usually characterized by several local and global measures [6]. The most important local measures are clustering coefficient $C$ and the average node degree $k$. Mathematically the clustering $C_i$ of the node $i$ is a probability, that the two neighbours of node $i$ are mutual neighbours as well. Network clustering coefficient $C$ represents an average of all $C_i$-s. Clustering coefficient is in fact a measure of nontrivial "structure" in the network. By non trivial is meant, that the network is not a tree or a simple regular lattice with nearest neighbour connections, only. As a global measure node separation $l$ (average shortest path between randomly chosen sites) is typically used. Separation of nodes shows, how "close" is one node to the other, or, in other words, how well the nodes communicate through edges.

Special type of network is a small world network [6]. It's structure optimises between the local regularity preservation, which tends to enhance node separation $l$ and good global node communication through random shortcuts. In this networks a high clustering coefficient $C$, is combined with a low node separation $l$.

As have been mentioned above, networks usually change their size with time [11, 13]. Many real networks, such as internet or word web, grow by the continual addition of new nodes. The node addition might be accompanied by node deletion, but the ratio of deleted nodes is often negligible. Therefore the dynamics of real networks is well captured by the models of growing nets [10, 12, 11, 13].

Many recent studies have shown [7, 1, 12, 14], that real networks, which are created by self organized processes, have common features. Their static properties are similar to that of small world nets. On the other hand, their degree distribution function, which is influenced by the dynamics, has power law character:

$$P(k) \propto k^{-\gamma}. \tag{1}$$

Such networks are called scale free [10]. The same properties has the word web [1, 3].

In this paper a positional word web is studied [1, 4]. Here the words are nodes and the word interaction is defined by the neighbourhood in a sentence. Language is a living phenomenon, developing all the time. Some words are created and some of them fall into disuse. Hence, to understand the word web dynamics, it is important to examine the dynamics of nets with changing number of sites.

This paper is organized as follows: In Section 2 the question of scale free network structure and dynamics is studied. Third Section is devoted to the mathematical models of positional word web and in the Section 4 my word web model is presented.

## 2  Scale free networks

As has been mentioned in the previous section, scale free structure is a result of self organized network development. Therefore this process should be natural and simple. The nature of the processes leading to the scale free structure has been investigated by Barabási and Albert in their fundamental paper [10]. In the Barabási - Albert model (BA model) the growth of net starts from small bunch of interconnected nodes. Each time unit a node comes and adjoins itself to the old network by $m$ new links. The probability of linking with certain old node $i$ is proportional to its degree $k_i$. Such type of linking is called preferential.

There are several possibilities, how do describe this processes mathematically. In many cases, the most efficient seems to be a continuous approach of Dorogovtsev and Mendes [13]. Newcoming nodes are labelled by their birth -time $s$. At time $t$, node $s$ has, in average, $k(s,t)$ neighbours. The average degree $k(s,t)$, is given by the equation

$$\frac{\partial k(s,t)}{\partial t} = m \frac{k(s,t)}{\int_0^t k(s,t)ds}. \tag{2}$$

Here the rhs of the equation (2) expresses how $k(s,t)$ changes by the preferential linking. To find the solution, the sum of all node degrees expressed as the renormalization integral $\int_0^t k(s,t)ds$ in the denominator, is to be estimated. It is easy. Each time unit $m$ new edges increase the sum by $2m$. Therefore

$$\int_0^t k(s,t)ds = 2mt. \qquad (3)$$

Substituting (3) into (2), equation (2) is easily solved [13]:

$$k(s,t) \propto \left(\frac{t}{s}\right)^\beta \propto s^{-\beta} \qquad (4)$$

where $\beta = \frac{1}{2}$. Having $k(s,t)$, power law degree distribution $P(k)$ (1) is easily analytically calculated [13]. But to get (1) together with the scaling exponent $\gamma$, such calculations are not necessary. It has been proven [13], that the exponents $\beta$ and $\gamma$ are related by the scaling relation

$$\gamma = 1 + \frac{1}{\beta}. \qquad (5)$$

From (5) and (4) one gets $\gamma_{BA} = 3$ and thus scale free degree distribution is

$$P(k) \propto k^{-3} \qquad (6)$$

in the BA model.

To summarize, preferential linking leads to the scale free network structure (6). Several other types of node linking were investigated. It has been shown, that if the node linking is a variation of preferential connection, the structure is scale free, but with $\gamma \neq \gamma_{BA}$ [11].

## 3   Positional word web

Lexicon of human language is composed of several ten thousands words. In spite of the huge amount of concepts, human brain is capable to manage them very quickly. Our speech is fluent, we are capable of quick retrieval in the large word database. How is it possible? How is human lexicon implemented in a brain? Of course, there are several theories about it. One of them says, that the lexicon has a structure of small world graph [6, 2, 1, 5].

Let us imagine a graph consisting of words as a nodes. Each word is connected by some edges (interactions) to the other words. It seems reasonable to define an interaction by the two different manners, which lead to the two different word nets, namely conceptual [2] and positional [1]. The first one is related to the semantics and the second one to the syntax. Both of them have small world properties, namely, large clustering combined with small node separation.

Positional net is related to the syntax and reflects the co-occurrence of words in a sentence. The words (graph sites) are connected by an edge, if they are neighbours in a sentence. In the human lexicon, two subsets of different size are recognized, namely the kernel lexicon, and the rest. Kernel lexicon includes about ten thousand most frequent words, known to the majority of people speaking the language. The other part, having hundred thousand words, is used in the various specialized communities. Studies of positional word web show, that its distribution function $P(k)$ scales as (1), but with two different exponents [4]. For well connected kernel words ($k$ is great), the scaling exponent is close to the theoretically predicted value $\gamma_{BA} = 3$. Less connected words scale with $\gamma = 1.5$. These two scaling regimes were explained by the model of Dorogovtsev and Mendes [4] (DM model).

The model is as follows: Each time unit a node comes and links itself preferentially by $m$ edges. Simultaneously $ct$ new edges (that means $2ct$ edge ends, $c << 1$) are created and connect the old nodes with preference. In this case $k(s,t)$ changes with time as:

$$\frac{\partial k(s,t)}{\partial t} = (m + 2ct)\frac{k(s,t)}{\int_0^t k(s,t)ds} \qquad (7)$$

where the integral gives the sum of node degrees

$$\int_0^t k(s,t)ds = 2mt + ct^2. \qquad (8)$$

With a help of (8) the solution of (7) is found [4]:

$$k(s,t) = \left(\frac{t}{s}\right)^{\frac{1}{2}} \left(\frac{2+ct}{2+cs}\right)^{\frac{3}{2}}. \qquad (9)$$

and two scaling regimes are recognised. For $s << t$ (well connected words) $\beta_{DM} = \frac{1}{2}$ and $\gamma_{DM} = 3$, and for $s \sim t$ (less connected words) $\beta_{DM} = \frac{1}{2} + \frac{3}{2}$ and $\gamma_{DM} = 1.5$ (5).

Let us check, how well the DM model describes measured data. The distribution $P(k)$ measured by Cancho and Solé [1], as well as our own studies $Fig.1$ [5] show, that there is a discrepancy between critical exponents predicted by the DM model and measured exponents. In the less steep part of the distribution $\gamma \approx 1.5$ and is the same, or very close to $\gamma_{DM}$. But $\gamma$ of the steeper part of the distribution shows the systematic error. In both cases it is lower (2.7 [1] and 2.13 [5]) then $\gamma_{BA} = 3$. I guess, that this is due to the fact, that the DG model doesn't include all processes significant for the word web. In the next section I propose a model which fit the data more accurately.

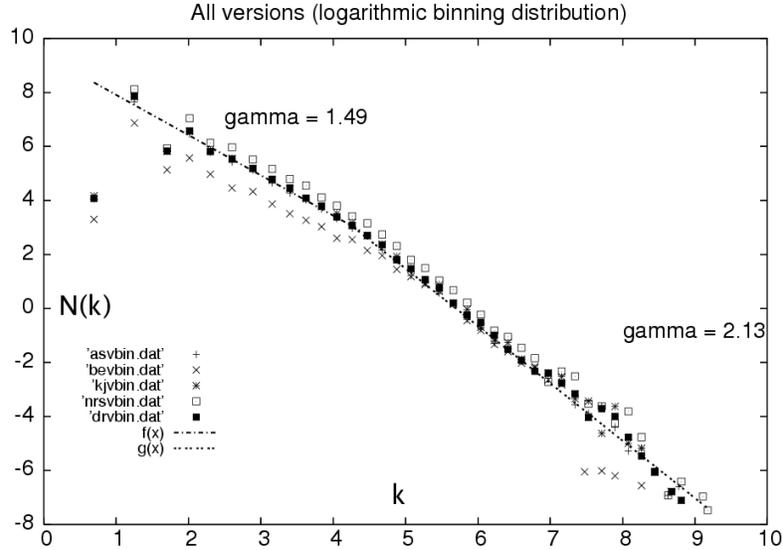

Figure 1: Connectivity distribution for the positional word web constructed of several English versions of The Bible (log -log plot). Some of them, such as King James version (kjv), Douay Rheims version (drv) are old (kjv has been issued in the year 1611, drv is even older, 1582), the others (American standard version, asv, 1901; Basic English versin, bev, 1941; New revisited standard version, nrsv, 1989) are relatively modern. bev is special, because its text has been artificially simplified. It is reflected in slightly different distribution.

## 4  Word web model

What are the other events, that should be considered in the positional word web? New words are created and added to the vocabulary all the time. They are used in sentences. Simultaneously old words might be used in a new phrases or contexts. In the word web this means a creation of new edges among old nodes. Both events are included in the DM model (7).

What are the other possibilities? As the time flows, the meaning of word might change slightly (or even significantly). In the word web some of the old connections are broken and rewired. Edge rewiring can be preferential, random, or a combination of both.

In [13] another model with preferential attachment is analytically solved: Every time unit a node is coming and linked preferentially to $m$ old nodes. In the same time $m_r = m_{r,p} + m_{r,r}$ old nodes are randomly selected. One edge end of $m_r$ nodes is rewired, $m_{r,p}$ of them are rewired with preference, and $m_{r,r}$ ends randomly. The model is solved and scaling exponent $\gamma$ is found:

$$\gamma = 2 + \frac{m - m_{r,p}}{m + m_{r,p}}. \tag{10}$$

If $m_{r,p} = 0$, $\gamma = \gamma_{BA}$. Because $m_{r,p} > 0$, rewiring lowers the $\gamma$ exponent and thus $\gamma < \gamma_{BA}$.

To fit the measured data [1, 5], ($Fig.1$) I designed a model which includes minimal amount of events. My model maintains two scaling regimes in the distribution function $P(k)$ (1). Likewise it explains, why the $\gamma$ exponent of the steeper part of $P(k)$ is below the BA value 3. In some sense, the model is a combination of a model with edge rewiring [13], and DM model [4].

Again, each time unit a node is added and linked preferentially with $m$ edges to the older nodes. Simultaneously another events occur:

1. $ct$ new edges are created and linked preferentially among old nodes;
2. $m_r$ old nodes are randomly selected and one edge end of them is rewired preferentially.

In the continuous approach these processes are described by the equation:

$$\frac{\partial k(s,t)}{\partial t} = (m + 2ct + m_r)\frac{k(s,t)}{\int_0^t k(s,t)ds} - \frac{m_r}{t} \tag{11}$$

To solve the equation (11), the integral $\int_0^t k(s,t)ds$ should be specified. As in the previous model, it is a sum of all degrees in the net. This sum is changed only by the new link creation; rewiring left it unaffected. The edge creation processes are :

a) Edge addition - $m$ new links come each time unit with a new node.

b) Appearance of new edges - $ct$ new links, or $2ct$ new link ends appear each time unit among old nodes.

The rewiring process is:

c) $m_r$ nodes are randomly selected. Each of them loose one edge end. This is expressed in the element $\frac{m_r}{t}$, where the number of nodes at time $t$ is proportional to $t$. Each of these ends is rewired preferentially.

Hence, the number of new edges, which appear in the network up to time $t$ is exactly the same as in the DM model (7). Therefore

$$\int_0^t k(u,t)du = 2mt + ct^2. \qquad (12)$$

Of course, it is easy to get (12) formally, by integrating both sides of (11)

$$\int_0^t ds \frac{\partial k(s,t)}{\partial t} = (m + 2ct + m_r) - \frac{m_r \int_0^t ds}{t} = m + 2ct \qquad (13)$$

and with a help of the expression

$$\frac{\partial}{\partial t} \int_0^t k(s,t)ds = k(t,t) + \int_0^t ds \frac{\partial k(s,t)}{\partial t} = m + m + 2ct, \qquad (14)$$

one identifies (12). Substituting (12) into (11) the equation is:

$$\frac{\partial k(s,t)}{\partial t} = (m + 2ct + m_r) \frac{k(s,t)}{2mt + ct^2} - \frac{m_r}{t} \qquad (15)$$

Because $m_r$ is a constant, for $t \to \infty$, $\frac{m_r}{t} \to 0$. Using this (15) is simplified and analytically solved

$$k(s,t) \propto \left(\frac{t}{s}\right)^{\frac{m+m_r}{2m}} \left(\frac{2m+ct}{2m+cs}\right)^{2-\frac{m+m_r}{2m}}. \qquad (16)$$

The solution of (11) is similar to that of (9), but with different $\beta$ exponents.

-if $s << t$, $\beta = \frac{m+m_r}{2m}$ and due to (5) $\gamma = 2 + \frac{m-m_r}{m+m_r}$,

-but if $s \sim t$, $\beta = 2 - \frac{m+m_r}{2m} + \frac{m+m_r}{2m} = 2$ and due to (5) $\gamma = 1.5$.

It is clear, that in my model the scaling exponent $\gamma$ is in the region of great $k$ lower then the value $\gamma_{BA} = 3$, but maintains the value 1.5 for the region of small $k$. This is exactly what was measured by Cancho and Solé and by us [1, 5], ($Fig.1$). The model (11) seems to fit the data better, then the former DM model [4].

Let us speculate a little. Our measurement shows [5],($Fig.1$), that $\gamma = 2.13$ for great $k$. Let us suppose, that newborn word has about ten connections $m \approx 10$. In that case the number of rewired edge ends is $m_r \approx 7.7$.

## 5  Conclusion

In conclusion, I present a model of growing network, which includes several local events, such as preferential link addition and preferential link rewiring. The model qualitatively and quantitatively correctly describes measured word web data. My model is inspired by DM model of growing network [4]. Additional local processes of edge rewiring cause, that the scaling exponent $\gamma$ of distribution function (1) is lower then the exponent of fully analytically solvable and well known BA model [10] (2). These local events are:

a) random node exclusion, and
b) preferential rewiring of one link end of the chosen node.

In our word web this processes mean, that certain word looses one of its meaning, or context, and another one is used in different context. For example, the word "notebook" has denoted exercise book for children. Now it is used more in a context of computers and informatics. Another example: "computer" in fifties was a big device. To tell anybody to put the computer on the table was nonsense. Now it is perfectly OK.

More detailed analysis of our data indicates, that the scaling exponent $\gamma$ might be slightly lower then 1.5 for small $k$. This is also supported by our analysis of another texts [15]. I therefore suppose, that there are another processes, such as node aging [13] or random edge rewiring [11] that might play some role. To investigate their relevance is a future task.

Acknowledgement: I acknowledge fruitfull discussions with Dr. Tomáš Blažek, prof. Pavol Brunovský, Dr. Michal Demetrian and prof. Ján Filo from the Faculty of Mathematics, Physics and Informatics, Comenius University, Bratislava, Slovakia, and prof. Peter Tiňo from the School of Computer Science, University of Birmingham, UK. This work has been supported by VEGA grants NO 1/2045/05, 2/4026/04, 1/3612/06 and 2/7087/27.